\documentclass{article}
\usepackage{isit2004}
\usepackage{pdfsync}
\usepackage{graphicx}
\usepackage{amssymb}
\usepackage{amsmath}

\begin{document}

\itwtitle{Performance of the Bounded Distance Decoder on the AWGN Channel}
 \itwauthorswithsameaddress {Kenneth Andrews and Sam Dolinar\footnotemark[1]}
           {Jet Propulsion Laboratory, California Institute of Technology, Pasadena, CA, USA\\
           e-mail: {\tt \{andrews,sam\}@shannon.jpl.nasa.gov}}

\itwmaketitle

\footnotetext[1]{\copyright \ 2012 California Institute of Technology.  Government sponsorship acknowledged.}

\begin{itwabstract}
In contrast to a maximum-likelihood decoder, it is often desirable to use an incomplete decoder that can detect its decoding errors with high probability.  One common choice is the bounded distance decoder.  Bounds are derived for the total word error rate, $P_w$, and the undetected error rate, $P_u$.  Excellent agreement is found with simulation results for a small code, and the bounds are shown to be tractable for a larger code.
\end{itwabstract}

\begin{itwpaper}

\itwsection{Introduction}
Error correcting codes are used in many settings where it is desirable to correct errors when it can be done sufficiently reliably, and to report a decoding failure when it cannot.  In spacecraft telecommand, for example, it is far preferable to discard a command than it is to deliver one that has a significant probability of being incorrect.

Three classes of incomplete decoders were considered in \cite{DAPD08a}: the optimal Bounded Reciprocal Likelihood Ratio (BRLR) decoder, the Bounded Angle (BA) decoder, and the Bounded Distance (BD) decoder.  The BA decoder was further studied in \cite{DAPD07,DAPD08b}, but the bounds are difficult to use.  While the BD decoder does not perform especially well, it is an important class of decoders, and practical performance bounds can be derived.

The ``amount of incompleteness'' of a BD decoder can be set by its decoding radius $r_d$.  It is commonly agreed that a BD decoder reports a decoding failure if no codeword lies within $r_d$ of the received noisy vector $\mathbf{y}$, and if a single codeword lies within $r_d$, the decoder returns that result.  If $r_d\geq d_{\min}/2$, more than one codeword may lie within $r_d$, and in this paper, we specify that the BD decoder returns the closest codeword.  Thus, the BD decoder asymptotically becomes the Maximum Likelihood (ML) decoder as $r_d$ is made large.

\itwsection{Performance Bounds}
\label{sec:bounds}
The BD decoder returns three outcomes: it finds the correct codeword with probability $P_c$, it makes an undetected error and returns an incorrect codeword with probability $P_u$, and it declares a decoding failure with probability $P_f$.  Here, we derive upper bounds for $P_u$ and for the total word error rate $P_w=P_u+P_f$, each as functions of $r_d$ and the bit Signal to Noise Ratio $E_b/N_0$.

\begin{figure}[tbh]
\begin{center}
\includegraphics[clip,trim = 75mm 110mm 45mm 85mm,width=3in]{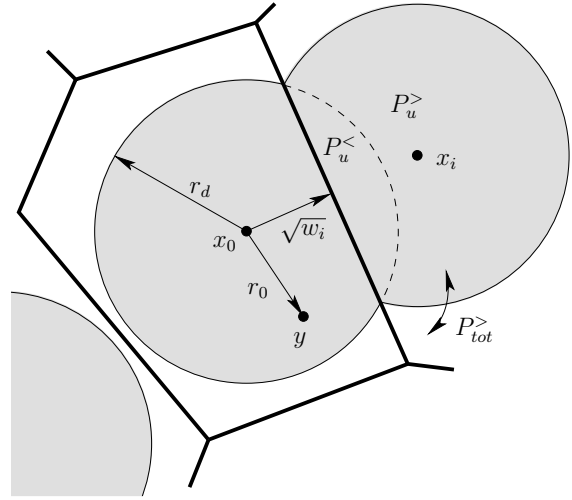}
\caption{Geometry of the Bounded Distance Decoder}
\label{fig:venn3}
\end{center}
\end{figure}

We consider an $(n,k)$ code with rate $R=k/n$.  Let the code have codewords $\mathbf{x}_i$, $0\leq i< 2^k$.  Suppose $\mathbf{x}_0$ is transmitted, and is received as $\mathbf{y}=\mathbf{x}_0+\mathbf{N}$, where each of the $n$ components of $\mathbf{N}$ are zero-mean Gaussian variables with variance $\sigma^2=1/(2RE_b/N_0)$.  The geometry of the BD decoder is shown in Figure \ref{fig:venn3}, where shaded spheres of radius $r_d$ are shown around codeword $x_0$ and another representative codeword $x_i$.  An undetected error occurs if $\mathbf{y}$ lands in a shaded region outside $x_0$'s Voronoi cell (shown in bold black lines), and a word error also occurs if $\mathbf{y}$ lands in an unshaded region.  That is,
\[ P_w=P_u^<(r_d)+P_{tot}^{>}(r_d) \]
\[ P_u=P_u^<(r_d)+P_u^>(r_d) \]
The constituent probabilities can be computed as integrals of the surface areas of caps of $n$-dimensional spheres (or $n$-sphere).  The surface area of an $n$-sphere of radius $r$ is
\[ S_n(r)=\frac{2\pi^{n/2}}{\Gamma(n/2)}r^{n-1}. \]
The surface area of the cap of an $n$-sphere with polar half-angle $\phi$ and radius $r$ is
\[ S_n(r,\phi)=\frac{1}{2}S_n(r)I_{\sin^2\phi}\left(\frac{n-1}{2},\frac{1}{2}\right) \]
where $I$ is the regularized incomplete beta function \cite{Li11}.

The probability density of receiving $\mathbf{y}$ when $\mathbf{x_0}$ is transmitted is
\[ p(\mathbf{y}|\mathbf{x_0})=\frac{1}{(\sigma\sqrt{2\pi})^n}e^{-||\mathbf{y}-\mathbf{x_0}||^2/2\sigma^2}. \]
Hence, the probability density that $||\mathbf{y}-\mathbf{x}_0||=r$ is
\begin{eqnarray*}p_0(r) & = & S_n(r)\sigma^{-n}(2\pi)^{-n/2}e^{-r^2/2\sigma^2} \\
& = & \frac{2^{1-n/2}}{\Gamma(n/2)}\sigma^{-n}r^{n-1}e^{-r^2/2\sigma^2} \\
& = & \frac{1}{\sigma}f\left(\frac{r}{\sigma},n\right) \end{eqnarray*}
where $f$ is the PDF of the $\chi$-distribution.

Now we can work out the component probabilities.
\begin{eqnarray*} P_{tot}^>(r_d) & = & \int_{r_d}^{\infty} p_0(r)\ dr \\
& = & 1-\frac{1}{\sigma}\int_0^{r_d} f\left(\frac{r}{\sigma},n\right)\ dr \\
& = & 1-F\left(\frac{r_d}{\sigma},n\right) \\
& = & Q\left(\frac{n}{2},\frac{r_d^2}{2\sigma^2}\right)
\end{eqnarray*}
where $F$ is the CDF of the $\chi$-distribution and $Q$ is the regularized Gamma function.

By a union bound over all incorrect codewords $x_i$, we have
\begin{equation} P_u^<(r_d) \leq \sum_{i\neq0} \mathrm{Pr}(r_i\leq r_0\leq r_d) \label{eq:Puless} \end{equation}
From Figure \ref{fig:venn3}, we have,
\[ \mathrm{Pr}(r_i\leq r_0\leq r_d)=\left\{\hspace{-3pt} \begin{array}{ll}
\displaystyle \int_{\sqrt{w_i}}^{r_d} p_0(r)\frac{S_n(r,\phi_i(r))}{S_n(r)}\ dr\hspace{-3pt} &\mbox{if } \sqrt{w_i}\leq r_d \\
0 & \mbox{otherwise} \end{array} \right. \]
where $S_n(r,\phi_i(r))$ is the surface area of the hyperspherical cap of radius $r$ and half-angle $\phi_i(r)$.  Note that $||\mathbf{x_0}-\mathbf{x_i}||/2=\sqrt{w_i}$, where $w_i$ is the Hamming weight of $\mathbf{x_i}$, and $\cos \phi_i(r)=\sqrt{w_i}/r$.
Substituting into (\ref{eq:Puless}),
\begin{eqnarray*} P_u^<(r_d) & \leq & \sum_{\substack{i\neq0 \\ \sqrt{w_i}\leq r_d}} \int_{\sqrt{w_i}}^{r_d} p_0(r) \frac{S_n(r,\phi_i(r))}{S_n(r)}\ dr \\
& = & \sum_{\substack{i\neq0 \\ w_i\leq r_d^2}}  \int_{\sqrt{w_i}}^{r_d} \frac{1}{2}p_0(r) I_{1-w_i/r^2}\left(\frac{n-1}{2},\frac{1}{2}\right)\ dr \\
& = & \sum_{w=d_{\min}}^{\min(n,r_d^2)} \frac{A_w}{2} \int_{\sqrt{w}}^{r_d} p_0(r) I_{1-w/r^2}\left(\frac{n-1}{2},\frac{1}{2}\right)\ dr
\end{eqnarray*}
where $A_d$ is the code's weight enumerator and $d_{\min}$ is the code's minimum distance.

Likewise, we can compute $P_u^>(r_d)$ from a union bound over all incorrect codewords $x_i$:
\begin{equation} P_u^>(r_d) \leq \sum_{i\neq 0} \mathrm{Pr}(r_i\leq r_d<r_0) \label{eq:Pugtr} \end{equation}
For each $i$,
\[ \mathrm{Pr}(r_i\leq r_d<r_0)=\int_{\max(r_d,2\sqrt{w_i}-r_d)}^{2\sqrt{w_i}+r_d} p_0(r)\frac{S_n(r,\phi_i(r))}{S_n(r)}\ dr \]
where in this case the half-angle of the spherical cap is given by the law of cosines:
\[ r_d^2=(2\sqrt{w_i})^2+r^2-2(2\sqrt{w_i})(r)\cos \phi_i(r) \]
Substituting into (\ref{eq:Pugtr}) we have,
\begin{eqnarray*}
\lefteqn{P_u^>(r_d) \leq \sum_{i\neq 0} \int_{\max(r_d,2\sqrt{w_i}-r_d)}^{2\sqrt{w_i}+r_d} p_0(r)\frac{S_n(r,\phi_i(r))}{S_n(r)}\ dr} \\
& & = \sum_{i\neq 0} \int_{\max(r_d,2\sqrt{w_i}-r_d)}^{2\sqrt{w_i}+r_d} \frac{1}{2}p_0(r)I_{\sin^2\phi_i(r)}\left(\frac{n-1}{2},\frac{1}{2}\right)\ dr \\
& &\hspace{-15pt} = \sum_{w=d_{\min}}^n \frac{A_w}{2} \int_{\max(r_d,2\sqrt{w}-r_d)}^{2\sqrt{w}+r_d} p_0(r) I_{\sin^2\phi_w(r)}\left(\frac{n-1}{2},\frac{1}{2}\right)\ dr
\end{eqnarray*}
where
\[ \sin^2 \phi_w(r)=1-\left(\frac{r^2-r_d^2+4w}{4r\sqrt{w}}\right)^2 \]

\itwsection{Examples}

The (8,4) extended Hamming code has $n=8$, $k=4$, $d_{\min}=4$, and weight distribution $A_w=\{1,0,0,0,14,0,0,0,1\}$ for $w=\{0, \ldots, 8\}$. The equations in Section \ref{sec:bounds} were evaluated for this code using Mathematica for $E_b/N_0=6.7$ dB (for which an ML decoder achieves $P_w\approx 10^{-4}$), and are plotted as curves in Figure \ref{fig:BoundedDistSims}.  Computer simulations were run for $10^{10}$ codewords, and the error rates are plotted as points.  The black curve for $P_{tot}^>$ is exact; the others are upper bounds that are evidently good except where $P_u^>$ is small.

\begin{figure}[tbh]
\begin{center}
\includegraphics[clip,trim = 0mm 0mm 0mm 0mm,width=3.5in]{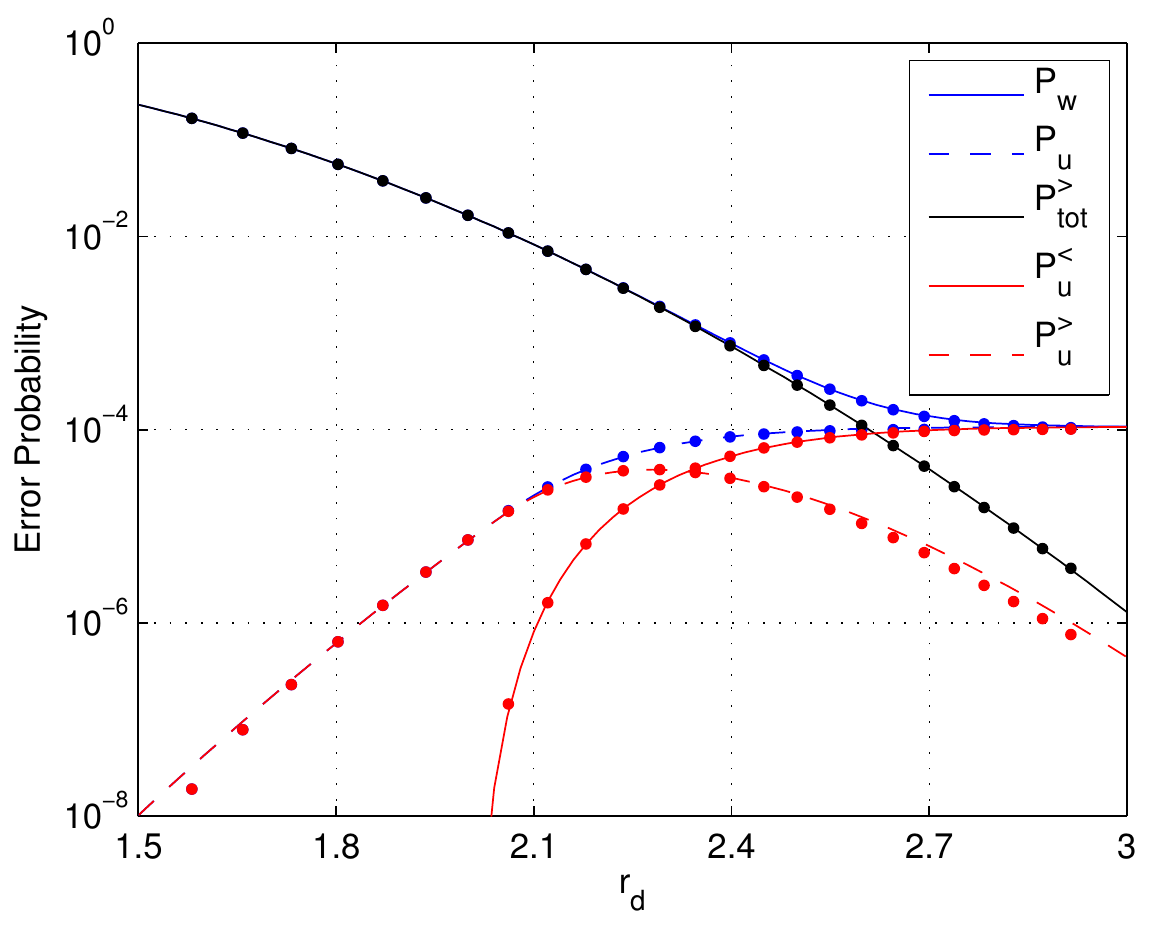}
\caption{Bounds and simulation results for the (8,4) extended Hamming code}
\label{fig:BoundedDistSims}
\end{center}
\end{figure}

A set of small rate-1/2 LDPC codes was developed for spacecraft telecommand \cite{DDJ07}.  For the $(128,64)$ code, computer search shows $d_{\min}=14$ and the weight enumerator is approximately $A_w=\{16,0,512,0,5344,0,\ldots\}$ for $w=\{14,15,\ldots\}$.  An LDPC decoder detects most of its errors, and its incompleteness can be adjusted by varying the number of iterations performed. While an LDPC decoder is not a BD decoder, its performance for this code is well predicted by the equations derived here.

\end{itwpaper}

\begin{itwreferences}
\bibitem{DAPD08a} Dolinar, Andrews, Pollara, Divsalar, ``The Limits of Coding with Joint Constraints on Detected and Undetected Error Rates'', ISIT (Toronto), July 6-11, 2008.
\bibitem{DAPD07} Dolinar, Andrews, Pollara, Divsalar, ``Bounded Angle Iterative Decoding of LDPC Codes'', Milcom, 2007.
\bibitem{DAPD08b} Dolinar, Andrews, Pollara, Divsalar, ``Bounds on Error Probability of Block Codes with Bounded-Angle Maximum-Likelihood Incomplete Decoding'', ISITA (Auckland), Dec. 7-10, 2008.
\bibitem{Li11} S. Li, ``Concise Formulas for the Area and Volume of a Hyperspherical Cap'', Asian Journal of Mathematics and Statistics, 2011.
\bibitem{DDJ07} Divsalar, Dolinar, Jones, ``Short Protograph-Based LDPC Codes'', Milcom, 29-31 Oct. 2007.
\end{itwreferences}

\end{document}